\documentclass[]{aa}
\usepackage{graphics,latexsym,psfig}
\usepackage{graphicx}

\def\farcs{\hbox{$.\!\!^{\prime\prime}$}}  % Fractions of arcseconds
\def\farcm{\hbox{$.\!\!^{\prime}$}}  % Fractions of arcminutes
\def\fdegs{\hbox{$.\!\!^{\circ}$}}  % Fractions of degrees
\def\asec{\ifmmode ^{\prime\prime}\else$^{\prime\prime}$\fi}
\def\amin{\ifmmode ^{\prime}\else$^{\prime}$\fi}
\def\degs{\ifmmode ^{\circ}\else$^{\circ}$\fi}
\def\etal{{et\,al. }}

\begin{document}

\title{Discovery of the Near-IR Afterglow and of the Host of GRB
       030528 \thanks{Based on observations collected at the European
       Southern Observatory, La Silla and Paranal, Chile by GRACE
       under ESO Program 71.D-0355.}  }

\titlerunning{}

\author{
A. Rau \inst{1}        \and
J. Greiner \inst{1}    \and 
S. Klose \inst{2}      \and 
M. Salvato\inst{1} \and
J.M. Castro Cer\'on\inst{3} \and
D.H. Hartmann\inst{4} \and
A. Fruchter \inst{3}   \and
A. Levan\inst{3}	\and
N.R. Tanvir\inst{5}  \and
J. Gorosabel\inst{3,6}   \and
J. Hjorth\inst{7}       \and
A. Zeh\inst{2} \and
A. K\"{u}pc\"{u} Yolda\c{s}\inst{1}    \and
J. P. Beaulieu\inst{8}   \and
J. Donatowicz\inst{9}    \and
C. Vinter\inst{7}        \and
A.J. Castro-Tirado\inst{6}   \and
J.P.U. Fynbo\inst{7,10}        \and
D.A. Kann\inst{2} \and
C. Kouveliotou \inst{11} \and
N. Masetti\inst{12}     \and 
P. M{\o}ller\inst{13}     \and
E. Palazzi \inst{12}    \and
E. Pian \inst{12,14}    \and 
J. Rhoads\inst{4}     \and  
R.A.M.J. Wijers \inst{15} \and
E.P.J. van den Heuvel \inst{15} 
}

\offprints{A. Rau, arau@mpe.mpg.de}

\institute{Max-Planck-Institut f\"ur extraterrestrische Physik,
  Giessenbachstrasse, 85748 Garching, Germany 
\and Th\"uringer
  Landessternwarte Tautenburg, 07778 Tautenburg, Germany 
\and Space Telescope Science Institute, 3700 San Martin Drive, Baltimore, MD
  21218, USA
\and Clemson University, Department of Physics and Astronomy, Clemson,
  SC 29634-0978, USA
\and Department of Physical Sciences, Univ. of Hertfordshire, College
  Lane, Hatfield Herts, AL10 9AB, UK 
\and Instituto de Astrof\'{\i}sica de Andaluc\'{\i}a (IAA-CSIC),
  Apartado de Correos, 3.004, E--18.080 Granada, Spain
\and Niels Bohr Institute, Astronomical Observatory, University of
  Copenhagen, Juliane Maries Vej 30, 2100 Copenhagen, Denmark 
\and Institut d'Astrophysique CNRS, 98bis Boulevard Arago, F--75014 Paris, France
\and Technical University of Vienna, Dept. of Computing, Wiedner Hauptstrasse 10, Vienna, Austria
\and Department of Physics and Astronomy, University of Aarhus, Ny Munkegade, 8000 Aarhus C, Denmark
\and NSSTC, SD-50, 320 Sparkman Drive, Huntsville, AL 35805, USA 
\and IASF/CNR, Sezione di Bologna, Via Gobetti 101, I--40129 Bologna, Italy 
\and European Southern Observatory, Karl Schwarzschild-Strasse 2,
  85748 Garching, Germany 
\and INAF, Osservatorio Astronomico di Trieste, Via Tiepolo 11, 34131 Trieste, Italy 
\and Astronomical Institute 'Anton Pannekoek', NL-1098 SJ Amsterdam, The Netherlands }

% --------------------------------------------------------------------------
\date{Received ?? 2004 / Accepted ?? 2004}

\abstract{The rapid dissemination of an arcmin-sized {\it HETE-2}
localization of the long-duration X-ray flash \object{GRB 030528} led to a
ground-based multi-observatory follow-up campaign. We report the
discovery of the near-IR afterglow, and also describe the detection of
the underlying host galaxy in the optical and near-IR bands. The
afterglow is classified as ``optically dark'' as it was not detected
in the optical band. The $K$-band photometry presented here suggests
that the lack of optical detection was simply the result of
observational limitations (lack of rapid and deep observations plus
high foreground extinction). Simple power law fits to the afterglow in
the $K$-band suggest a typically decay with a slope of
$\alpha$=1.2. The properties of the host are consistent with the idea
that GRB hosts are star forming blue galaxies. The redshift of
GRB 030528 can not be determined accurately, but the data favour
redshifts less than unity. In addition, we present an optical and
near-IR analysis of the X-ray source \object{CXOU J170354.0--223654} from the
vicinity of GRB 030528.
}
\maketitle

% --------------------------------------------------------------------------
\section{Introduction}

Optical and near-IR afterglows play a crucial role in the
understanding of the phenomenon of long duration Gamma-ray bursts
(GRB).  While prompt $\gamma$-ray emission has been known since 1973
(Klebesadel \etal 1973), a major breakthrough in GRB research came
with the discoveries of the first X-ray afterglow (Costa \etal 1997)
and optical transient (van Paradijs \etal 1997). Firm evidence for the
cosmological origin of GRBs was first obtained with the determination
of the redshift of $z$=0.835 for GRB 970508 from absorption lines in
the optical afterglow (Metzger \etal 1997). To date, afterglows for 75
well localized long duration GRBs have been detected and 36 redshifts
from emission lines in the underlying host galaxy and/or absorption
features in the optical afterglow were determined (see J. Greiner's
web page\footnote{http://www.mpe.mpg.de/$\sim$jcg/grbgen.html}). For
nearly all well localized bursts an X-ray afterglow was found whenever
X-ray observations were performed, but only 53 bursts were also
detected in the optical and/or near-IR band. One day after the GRB,
optical transients exhibit $R$-band magnitudes that are typically in
the range of $\sim$19--22 and $K_s$-band magnitudes of
16--19. Optical/near-IR afterglow light curves can be characterized by
a power law in time, F$\propto$t$^{-\alpha}$, with $\alpha$$\sim$1.3
(van Paradijs \etal 2000). For the remaining 23 GRBs with X-ray and/or
radio afterglow no optical and/or near-IR transient could be
detected. For this group of bursts the term ``dark burst'' was
introduced. GRBs detected in the near-IR but lacking an optical
afterglow constitute a sub-group, and can be labeled ``optically dark
bursts''.

In many cases observational limitations can account for the
non-detection in the optical or near-IR-band. A slow reaction time, a
location in a crowded field, possibly high Galactic foreground
extinction, or unfavorable observing conditions, like bright moon and
twilight, can explain non-detections of the counterparts. {\it HETE-2}
revealed that rapid and accurate localizations of the prompt emission
in nearly all cases lead to the detection of an optical transient
(Lamb \etal 2004). However, this does not provide a valid explanation
for all dark bursts (Klose \etal 2003). In some cases , e.g. \object{GRB
970828} (Groot \etal 1998; Djorgovski \etal 2001) and \object{GRB 990506}
(Taylor \etal 2000) even rapid (less than half a day after the GRB)
and deep ($R$$>$23) observations did not reveal an afterglow, despite
a clearly fading source in the X-ray and/or radio band.

There are many reasons for the non-detection of the optical transients
of bursts with known X-ray or radio afterglows (e.g. Fynbo \etal
2001a; Lazzati \etal 2002). In addition to the observational biases
mentioned above, the existence of ``dark bursts'' may reflect a broad
distribution of physical parameters of the GRB itself or of its
environment, as in the case of GRB 990506 (see Taylor \etal 2000).
The rapidly decaying radio afterglow of this burst together with the
non-detection in the optical could be due to an extremely low-density
medium surrounding the GRB.

Since the spectroscopic confirmation of \object{SN2003dh} underlying the
afterglow of \object{GRB 030329} (Hjorth \etal 2003; Stanek \etal 2003) it is
now widely believed that long-duration GRBs are associated with the
death of massive stars (e.g., Heger \etal 2003). Because of the
short lifetime of these progenitors of $\sim$10$^6$\,years, they do
not propagate far from their birth place in star forming
regions. Consequently, the optical and near-IR emission could suffer
from significant attenuation in the dusty medium. The X-ray and radio
afterglow emission may still be observable. Whether a burst is
``dark'' or has a detectable optical/near-IR transient would
therefore depend on the conditions of the ISM in the vicinity of the
GRB. However, it is conceivable that dust destruction by the prompt
emission and early afterglow phase alters the circumstances (Galama \&
Wijers 2001; Galama \etal 2003).

Another possibility to explain ``dark bursts'' is to place them at
high redshift (Lamb \& Reichart 2000). The observed redshift
distribution of GRBs is very broad and currently ranges between
$z$=0.0085 to $z$=4.5 with a broad peak around $z\sim$1
(e.g. Jakobsson \etal 2004). GRBs at still higher redshifts are
expected based on the association with massive stars discussed
above. However, the sensitivity of stellar mass loss to metalicity
combined with the requirement that jets must successfully emerge from
the stellar envelope suggests that single, massive stars in the early
universe may not result in observable GRBs (e.g., Heger \etal 2003).
Alternative scenarios, some perhaps including binary stars, may very
well produce GRBs at redshifts above $z$=6, where the Lyman alpha
absorption edge will be shifted through the optical into the near-IR
band. The resulting Lyman alpha suppression could then easily account
for the lack of optical detections. On the other hand, observations
show that the high-$z$ explanation for ``darkness'' can not apply in
all cases. For example, GRB 970828 and \object{GRB 000210}
revealed underlying host galaxies at positions coincident with those
of the X-ray and radio afterglows (Djorgovski \etal 2001; Piro \etal
2002) for which spectra indicate redshifts of $z$=0.958 and
$z$=0.8463, respectively.

Here we report on the discovery of the near-IR afterglow of the
optically dark GRB 030528 and it's underlying host galaxy. After
describing the prompt emission properties and afterglow searches by
other teams (Sect. 2), we present our optical and near-IR observations
(Sect. 3) and their reduction (Sect. 4). We show the properties of the near-IR
afterglow and of the host galaxy (Sect. 5) and discuss the results in the
context of dark bursts (Sect. 6).

\section{GRB 030528}
\begin{figure}[h]
   \centering
   \includegraphics[width=0.5\textwidth,angle=0]{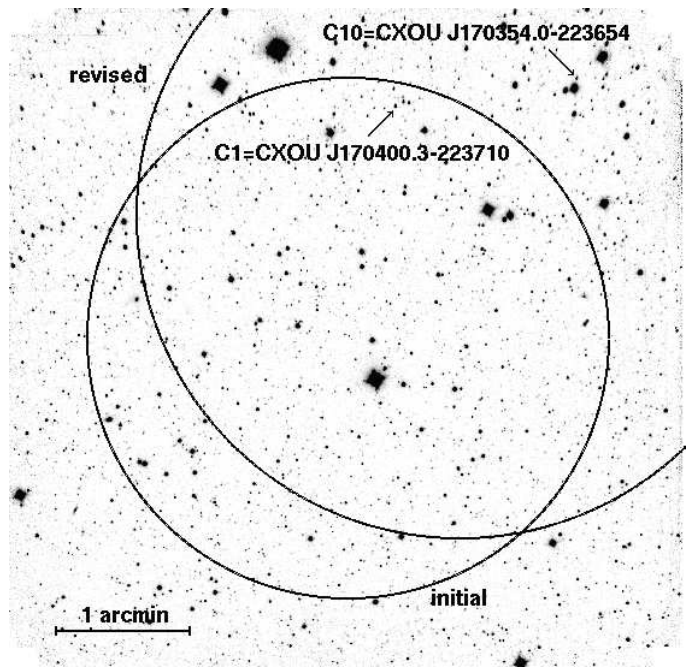} 
   \caption{15\,min exposure $K_s$-band SofI image obtained on May 29,
   2003. The initial and revised {\it HETE-2} SXC confidence circles are
   indicated together with the positions of two X-ray sources detected
   by {\it Chandra} (see text). North is up and East to the left.}
   \label{fig:fc} 
   \end{figure}

On May 28, 2003 the {\it HETE-2} French Gamma-ray Telescope (FREGATE)
and the Wide-field X-ray Monitor (WXM) triggered on a long-duration
Gamma-ray burst (HETE trigger \#2724) at 13:03 UTC. The event was
moderately bright with a
fluence\footnote{http://space.mit.edu/HETE/Bursts/Data} of
$S$=5.6$\times$10$^{-6}$ erg cm$^{-2}$ and a peak flux on a one second
time scale of 4.9$\times$10$^{-8}$ erg cm$^{-2}$ s$^{-1}$ in the
30-400\,keV band (Atteia \etal 2003a). The burst duration (given
as T$_{90}$, which is the time over which a burst emits from 5\% of
its total measured counts to 95\%, was T$_{90}$=49.1\,s (30-400\,keV)
and the high energy spectrum peaked at 32\,keV. The burst is
classified as an X-ray flash according to the fluence ratio $S_{(2-30
keV)}/S_{(30-400 keV)}$=1.13$\pm$0.15.The properties of XRFs, X-ray
rich bursts and GRBs apparently form a continuum (Lamb \etal 2004). At
107\,min after the onset of the burst a confidence circle with 2\amin\
radius centered at RA(J2000)=17h04m02s,
DEC(J2000)=$-$22\degs38\amin59\asec\ derived from the HETE-2 Soft
X-ray Camera (SXC) was released to the community. Figure~\ref{fig:fc}
shows a $K_s$-band finding chart centered on this initial error
circle.

%------------------------------------------------------------------
%------------------------------------------------------------------

Despite several rapid response observations, no optical afterglow was
found in this crowded field in the Galactic Plane (LII=0\fdegs0249,
BII=11\fdegs267). Table~\ref{tab:optImaging} provides a list of upper
limits. Later, a correction of the initial {\it HETE-2} error
circle had to be applied. The radius of the confidence circle increased
to 2\farcm5 and the centroid was displaced by 1\farcm3. This
modification was announced on May 31, 2.4\,days after the burst
(Villasenor \etal 2003; Fig.~\ref{fig:fc}).

%------------------------------------------------------------------
\begin{table}[h]
\begin{center}
\caption{Time after the burst, filters and limiting magnitudes of
  published optical/near-IR observations for the afterglow of GRB
  030528.}
\begin{tabular}{cccl}
\hline\hline
t & Filter & lim. mag & Ref. \\
\hline
106\,s & white & 15.8 & Torii 2003 \\
252\,s & white & 16.0 & Uemura \etal 2003 \\
0.097\,d & $R$ & 18.7 & Ayani \& Yamaoka 2003 \\
0.496\,d & white & 20.5 & Valentini \etal 2003 \\
5.831\,d & $K_s$ & 19.5 & Bogosavljevic \etal 2003 \\ 
7.653\,d & $I$ & 21.5 & Mirabal \& Halpern 2003 \\
\hline\hline
\end{tabular}
\label{tab:optImaging}
\end{center}
\end{table}
%------------------------------------------------------------------

A 26.1\,ksec {\it Chandra} observation performed on June 3
(5.97--6.29\,days after the burst) of the revised {\it HETE-2}
confidence circle revealed several X-ray sources in the 0.5--8\,keV
band (Butler \etal 2003a). Two of these, C1=CXOU J170400.3--223710 and
C10=CXOU J170354.0--223654 were located inside our field of
view. Following these detections, we inspected earlier multi-epoch
near-IR SofI images and found that one source, coincident with C1,
exhibited significant fading (Greiner et al. 2003a) which made this
source the most likely afterglow candidate. A second {\it Chandra}
observation (20\,ksec exposure on June 9) showed that only one of the
X-ray sources, C1, was in fact fading (Butler \etal 2003b). This
observation confirmed the identification of C1 as the afterglow of GRB
030528. The X-ray observations are described by Butler \etal
(2004). C1 is inside the initial {\it HETE-2} error circle. Therefore,
the optical non-detections mentioned above, are not a result of it's
later revision. Follow-up observations in the radio (Frail \& Berger
2003) did not detect C1, or any other source inside the revised error
circle.

% --------------------------------------------------------------------------
\section{Observations}

Shortly (0.6684\,days) after the initial 2\amin\ confidence circle was
released, ToO imaging with the 3.58\,m ESO-New Technology Telescope
(NTT) equipped with the Son of ISAAC (SofI) infrared spectrograph and
imaging camera at La Silla/Chile were initiated
(Tab.~\ref{tab:log}). SofI is equipped with a 1024$\times$1024 HgCdTe
Hawaii array with 18.\,5$\mu$m pixel size and a plate scale of
0\farcs29 per pixel. It has a field of view of 5\farcm5. During the
first two nights ($\sim$0.7 and $\sim$1.7\,days after the burst)
imaging in $J$, $H$ and $K_s$ was performed. During the fourth night
($\sim$3.6\,days post-burst) only $K_s$-band imaging was carried out. The
seeing conditions during the observations of this very crowded field
were good for the first and third epoch ($\sim$0\farcs8) but less
favorable for the second epoch (1\farcs1--1\farcs6). All of the above
imaging was centered on the initial {\it HETE-2} error circle. Due to
the later increase and shift of the confidence circle and the 5\farcm5
field of view of NTT-SofI, the observations do not cover the entire
revised error circle (see Fig.~\ref{fig:fc}).

At $\sim$14.9\,days after the burst, one $K$-band observation was
performed with the 3.8\,m United Kingdom Infra-Red Telescope
Fast-Track Imager (UKIRT UFTI) on Mauna Kea under good seeing
conditions (0.6\asec). UFTI consists of a 1024x1024 HgCdTe Rockwell
array with 18.5\,$\mu$m pixels and a plate scale of 0\farcs09 per
pixel, giving a field of view of 92\asec$\times$92\asec.

Nearly-Mould $I$-band photometry was obtained with the Mosaic2 imager
at the 4\,m Blanco Telescope at the Cerro Tololo Inter-American
Observatory (CTIO) 6.6 and 32.6\,days after the burst. The Mosaic2
consists of eight 2048x4096 SITe CCDs with a pixel size of
15\,$\mu$m. The plate scale of 0\farcs27 per pixel at the 4\,m
Blanco-Telescope produces a field of view of 36\amin$\times$36\amin.

In addition, late time $J_s$-band imaging was performed with the
Infrared Spectrometer And Array Camera (ISAAC) at the 8.2\,m ESO Very
Large Telescope (VLT) Antu in Paranal/Chile 111, 121, 124, and
125\,days after the burst. ISAAC is equipped with a 1024$\times$1024
pixel Rockwell Hawaii HgCdTe array with a 18.5\,$\mu$m pixel size. The
plate scale of 0\farcs147 per pixel provides a
2\farcm5$\times$2\farcm5 field of view. 

Further, late epoch $V$ and $R$-band observations were obtained with
the Danish Faint Object Spectrograph and Camera (DFOSC) at the Danish
1.54m Telescope at La Silla/Chile 381--386\,days after the
burst. DFOSC consists of a 2048$\times$4096 EEV/MAT CCD with a pixel
size of 15\,$\mu$m and a plate scale of 0\farcs39 per pixel. As the
instument optics does not utilise the full chip area, only
2148$\times$2102 pixels are illuminated, giving a field of view of
13\farcm$3\times$13\farcm3. These and the above mentioned observations
are summarized in Tab.~\ref{tab:log}.

\begin{table*}[h]
\begin{center}
\caption{Observation log. $<dt>$ stands for mid-observation time after
the burst. Magnitudes and flux densities are corrected for Galactic
foreground extinction. $^{(1)}$: all ISAAC $J_s$ observations combined.}
\begin{tabular}{crcccccc}
\hline\hline
Date & $<dt>$ & Telescope/Instrument & Filter & Exposure & Seeing &
Brightness & log Flux Density\\
(Start UT) & (days) & & & (min) & & (mag) & (erg/cm$^2$/s/\AA)\\
\hline
2003 May 29 04:58 & 0.6684 & NTT-SofI & J & 15 & 0.8\asec & 20.6$\pm$0.3 & --17.73$\pm$0.12 \\
2003 May 29 05:16 & 0.6809 & NTT-SofI & H & 15 & 0.8\asec & 20.3$\pm$0.4 & --18.05$\pm$0.16 \\
2003 May 29 05:32 & 0.6920 & NTT-SofI & K$_s$ & 15 & 0.8\asec & 18.6$\pm$0.2 & --17.83$\pm$0.08 \\
2003 May 30 04:54 & 1.6673 & NTT-SofI & J & 20 & 1.6\asec & $>$20.2 & $<$--17.57\\
2003 May 30 05:16 & 1.6826 & NTT-SofI & H & 20 & 1.1\asec & $>$19.9 & $<$--17.89 \\
2003 May 30 05:40 & 1.6993 & NTT-SofI & K$_s$ & 20 & 1.1\asec & 18.9$\pm$0.3 & --17.95$\pm$0.12 \\
2003 Jun 01 04:07 & 3.6486 & NTT-SofI & K$_s$ & 60 & 0.8\asec & 19.6$\pm$0.5 & --18.23$\pm$0.20 \\
2003 Jun 04 02:10 & 6.5604 & Blanco-Mosaic2 & I & 40 & 1.2\asec & 21.4$\pm$0.3 & --17.63$\pm$0.12 \\
2003 Jun 12 08:55 & 14.8680 & UKIRT-UFTI & K & 116 & 0.6\asec & 19.6$\pm$0.1 & --18.23$\pm$0.04\\
2003 Jun 30 03:00 & 32.5951 & Blanco-Mosaic2 & I & 40 & 1.1\asec & 21.2$\pm$0.3 & --17.55$\pm$0.12 \\
2003 Sep 17 00:12 & 111.4820 & VLT-ISAAC & J$_s$ & 50 & 0.6\asec & 21.0$\pm$0.2 & --17.89$\pm$0.08 \\
2003 Sep 27 00:27 & 121.4966 & VLT-ISAAC & J$_s$ & 60 & 0.9\asec & 21.1$\pm$0.3 & --17.93$\pm$0.12 \\
2003 Sep 29 23:38 & 124.4743 & VLT-ISAAC & J$_s$ & 94 & 0.7\asec & 20.7$\pm$0.1 & --17.77$\pm$0.04 \\
2003 Oct 01 00:04 & 125.4799 & VLT-ISAAC & J$_s$ & 60 & 0.6\asec & 20.7$\pm$0.2 & --17.77$\pm$0.08 \\
2004 Jun 15--18 & 382.5 & D1m54-DFOSC & R & 75 & 1.4\asec & 22.0$\pm$0.2 & --17.55$\pm$0.08\\ 
2004 Jun 18--19 & 384.5 & D1m54-DFOSC & V & 135 & 1.5\asec & 21.9$\pm$0.2 & --17.19$\pm$0.08\\ 
\hline
2003 Sep 17 -- Oct 01 & & ISAAC-combined$^{(1)}$ & J$_s$ & 264 & & 20.8$\pm$0.1 & --17.81$\pm$0.04 \\
\hline\hline
\end{tabular}
\label{tab:log}
\end{center}
\end{table*}

\section{Data Reduction}
\begin{figure}[h]
   \centering
   \includegraphics[width=0.49\textwidth,angle=0]{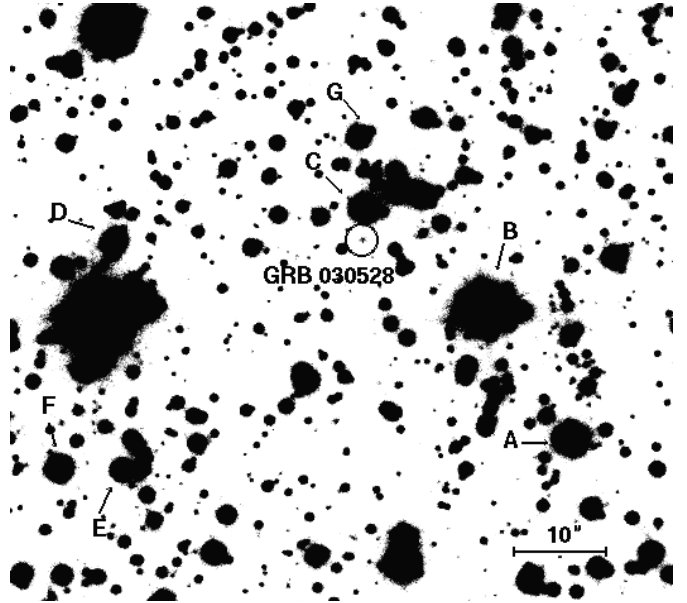} 
   \caption{Combined late time VLT/ISAAC $J_s$-band image from
   September 2003 with a total exposure of 264\,min. The position of
   GRB 030528 is marked by the circle, while the letters (from A-G)
   label the 2MASS stars used for the flux calibration (see also
   Tab.~\ref{tab:calibration}). The image size is
   $\sim$75\asec$\times$75\asec and North is up and East to the left.}
   \label{fig:chandra1Fc} 
   \end{figure}

The NTT-SofI and VLT-ISAAC near-infrared images were reduced using
ESO's {\it Eclipse} package (Devillard 1997). The reduction of the
Blanco-Mosaic2 data was performed with {\it bbpipe}, a script based on
the {\it IRAF/MSCRED}, the UKIRT-UFTI observation was reduced using
{\it ORACdr} and the DFOSC data were reduced with {\it
IRAF}. Astrometry was performed using {\it IRAF/IMCOORDS} and the
coordinates of stars in the field provided by the 2MASS All-Sky Point
Source
Catalog\footnote{http://irsa.ipac.caltech.edu/applications/Gator/}. For
the photometry we used {\it IRAF/DAOPHOT}. To account for the
distortions in the SofI images caused by the position of C1 at the
edge of the field of view (see Fig.~\ref{fig:fc}), we used stars
contained in the 2MASS Catalog from the vicinity of the source for the
photometric calibrations of the $J, J_s, H, K$ and $K_s$ fields.  The
stars are listed in Tab.~\ref{tab:calibration}. We only used the stars
for which magnitude uncertainties in the relevant bands were
provided. The Mosaic2 $I$-band images were calibrated using the USNOFS
field photometry of Henden (2003), in particular the three stars B, C
\& G (Fig.~\ref{fig:chandra1Fc}). The photometric measurements are
partly hampered by the combination of instrumental distortions and the
high density of sources in the field. The crowdedness of the field
also affected the set of comparison stars. Only source F
(Fig.\ref{fig:chandra1Fc}) is sufficiently isolated to provided high
quality calibration. The photometry for A, B, C, D, E and G is less
accurate in comparison to F, but not by much. In any case all
individual uncertainties are taken into account in the error
analysis. The DFOSC $V$ and $R$-band calibration was performed using
observations of the standard star G153-41 (Landolt 1992).

%------------------------------------------------------------------
\begin{table*}[h]
\begin{center}
\caption{Stars used for the flux calibration of the imaging data for
  C1 (Fig.~\ref{fig:chandra1Fc}). $I$-band magnitudes are from field
  photometry provided by Henden (2003) and $J$, $H$ \& $K_s$-band
  magnitudes are taken from the 2MASS All-Sky Point Source
  Catalog. $^{(1)}$: no $I$-band magnitudes available, $^{(2)}$: no
  uncertainties provided by 2MASS.}
\begin{tabular}{rcccccc}
\hline\hline
Star & RA (J2000) & DEC (J2000)& I & J & H & K$_s$ \\
& hh:mm:ss & dd:mm:ss & [mag] & [mag] & [mag] & [mag] \\
\hline
A & 17:03:58.6 & --22:37:33 & $^{(1)}$ & 14.97$\pm$0.03 & 14.26$\pm$0.05 & 14.04$\pm$0.05 \\
B & 17:03:59.3 & --22:37:18 & 14.60$\pm$0.02 & 13.37$\pm$0.02 & 12.64$\pm$0.03 & 12.46$\pm$0.02 \\
C & 17:04:00.3 & --22:37:06 & 16.33$\pm$0.10 & 15.50$\pm$0.07 & 14.87$\pm$0.08 & 14.76$\pm$0.11 \\
D & 17:04:02.5 & --22:37:10 & $^{(1)}$ & 16.09$\pm$0.11 & 15.58$\pm$0.11 & 15.48$\pm$0.21 \\
E & 17:04:02.9 & --22:37:37 & $^{(1)}$ & 16.57$\pm$0.13 & 16.15$\pm$0.17 & 15.76$^{(2)}$ \\
F & 17:04:02.4 & --22:37:37 & $^{(1)}$ & 15.96$\pm$0.08 & 15.48$\pm$0.10 & 15.20$\pm$0.18 \\ 
G & 17:04:00.4 & --22:36:58 & 17.44$\pm$0.20 & 16.43$\pm$0.11 &
15.88$\pm$0.12 & 15.38$^{(2)}$ \\
\hline\hline
\end{tabular}
\label{tab:calibration}
\end{center}
\end{table*}
%------------------------------------------------------------------

%------------------------------------------------------------------
%------------------------------------------------------------------

To establish the proper photometric zero-points for the different
instruments used in this study we cross-checked stellar colors (of
non-saturated stars in the field) against a set of theoretical colors
along the main sequence. We utilized synthetic stellar spectra from
the library of Pickles (1998) and convolved those with the filter
transmission curves and efficiencies of SofI, ISAAC, Mosaic2 and
DFOSC. A good match to within a zero-point accuracy of $\pm$0.05\,mag
is obtained in all bands.

The 2MASS catalog provides the standard $J$, $H$ \& $K_s$-band
magnitudes. In addition to these bands we also present observations
obtained in $J_s$ and $K$. The $J_s$-band filter has a width of
0.16\,$\mu$m and is narrower than $J$-band filter (0.29\,$\mu$m) and
the $K$-band filter is broader than the $K_s$-band filter with a width
of 0.35\,$\mu$m (instead of 0.27\,$\mu$m). Furthermore, the $K_s$-band
is centered at 2.16\,$\mu$m while the $K$-band is centered at
2.20\,$\mu$m. However, since $J_s$ and $K$ have respectively higher
and lower transmission than the $J$ and $K_s$, the net effect is that
$J$-$J_s\leq$0.05 ($K$-$K_s\leq$0.02).

All magnitudes are corrected for Galactic foreground extinction
according to the prescription given by Schlegel \etal (1998). For the
coordinates of the afterglow of GRB 030528 we find $E(B-V)$=0.60,
A$_K$=0.22\,mag, A$_H$=0.35\,mag, A$_J$=0.54\,mag, A$_I$=1.17\,mag,
A$_R$=1.61\,mag and A$_V$=2.00\,mag.

\section{Results}

We now describe our results on the near-IR afterglow and on the
optical/near-IR observations of the underlying host galaxy of GRB
030528 and discuss host properties in terms of population synthesis
models. We apply the same methodology to the source C10, which is
unrelated to the GRB, but defer the description to Appendix A.

C1 is the brightest X-ray source in the initial {\it Chandra} field at
a flux level of 1.4$\times$10$^{-14}$ erg cm$^{-2}$ s$^{-1}$ at
0.5--8\,keV (Butler \etal 2004). This value was calculated assuming a
power law spectrum with a slope of $\Gamma$=1.9 and taking into
account Galactic foreground extinction due to a neutral hydrogen
column density of 1.6$\times$10$^{21}$ cm$^{-2}$. At the X-ray
position of C1 a faint object in our SofI $J$, $H$ \& $K_s$-band
observations $\sim$0.7\,days post-burst is apparent. The source is
near the detection limits in $J$ and $H$, but significantly detected
in $K_s$. The magnitudes, corrected for foreground extinction in the
Galaxy, are $J$=20.6$\pm$0.3, $H$=20.3$\pm$0.4 and $K_s$=18.6$\pm$0.2
(see also Tab.\ref{tab:log}).

\begin{figure}
  \includegraphics[height=.4\textheight,bbllx=1.1cm,bblly=2.5cm,bburx=20.5cm,bbury=29.8cm,clip=,angle=-90]{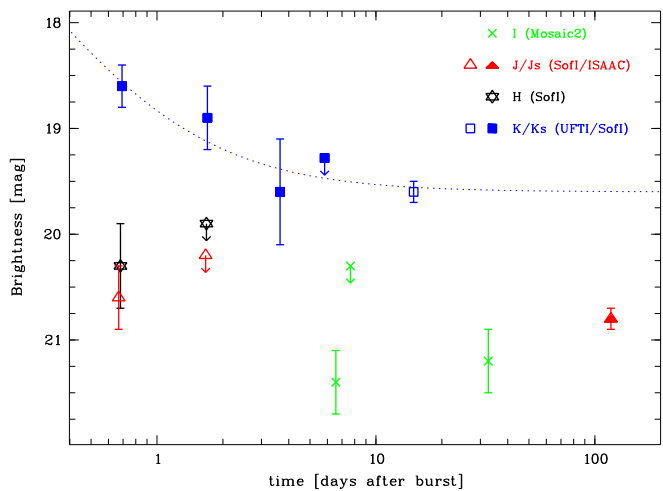}
  \caption{Foreground extinction corrected magnitudes in $I$ (marked
  by a cross), $J$/$J_s$ (triangle/filled triangle), $H$ (star) and
  $K$/$K_s$ (square/filled square). For $J$ and $H$ only upper limits
  exist for t$\sim$1.7\,days. The $K_s$-band upper limit at
  t=5.8\,days is from observations with the WIRC at the Palomar
  200-inch Hale telescope (Bogosavljevic \etal 2003) and the $I$-band
  upper limit at t=7.6\,days from the 1.3\,m McGraw-Hill telescope at
  the MDM observatory (Mirabal \& Halpern 2003). The dotted line
  corresponds to a decay with a power law slope of $\alpha$=1.2 with
  the host magnitude fixed to $K$=19.6.}
  \label{fig:lightcurves}
\end{figure}

Seeing conditions during the second night ($t\sim$1.6\,days
post-burst) only allow us to derive brightness limits for C1 in the
$J$ and $H$ band but we were able to detect the source at
$K_s$=18.9$\pm$0.3. The $J$ and $H$-band data are insufficient to test
variability, and the two $K_s$-band measurements are formally
consistent with a constant source. However, on June 1 (3.6\,days
post-burst) fading became apparent. At that time the source had
declined to $K_s$=19.6$\pm$0.5, corresponding to a change by roughly
$\sim$1\,mag within three days. Fig.~\ref{fig:lightcurves} shows the
light curve of the source in $K_s$ together with all near-IR
observations presented here and the near-IR upper limits published in
the GRB Coordinates Network
(GCN)\footnote{http://gcn.gsfc.nasa.gov/gcn/gcn3\_archive.html}.

%------------------------------------------------------------------
%------------------------------------------------------------------

In contrast to the $K$-band variability, no fading is observed in the
$I$-band observations taken 6.6 and 32.6\,days after the burst. The
source is persistent at a brightness of $I$=21.3$\pm$0.3, which we
interpret as the $I$-band magnitude of the host galaxy. Similarly,
comparing the late time ($>$100\,days post-burst) ISAAC $J_s$-band
observations with the SofI $J$-band data from the first night
($t\sim$0.6\,days post-burst), the source also remains constant within
the uncertainties of the measurements. Thus, the decay of the near-IR
afterglow is only detected in the $K_s$-band.

In order to compare the afterglow decay in the near-IR with that in
the X-ray band, we estimate the power law slope, $\alpha$, from the
few $K/K_s$-band data shown in Fig.~\ref{fig:lightcurves}. Obviously,
the uncertainties in the photometry and the poor sampling of the light
curve do not allow us to derive an accurate description of the
afterglow behavior. A major source of uncertainty is introduced by the
fact that we do not know when a jet break may have occurred. If the
break occurred before our first $K_s$-band observation at t=0.7\,days
the subsequent decay slope may have been close to $\alpha$=1.2. If the
afterglow is best described by a single power law, the shallowest
slope could be around $\alpha$=0.7. In that case, the afterglow
contribution to the $K$-band flux at 14.9\,days post-burst is not
negligible. Slopes much steeper than $\alpha$=1.2 can be imagined
upon arbitrarily placing the break time close to t=3\,days. This is
likely to be the case considering typical break times of
$t$$\sim$0.4--4\,days. Therefore, it appears reasonable that the
post-break near-IR slope falls in the range of $\alpha$=0.7--2. The
significant uncertainties in the slopes leave the possibility that the
near-IR and X-ray ($\alpha$=2.0$\pm$0.8; Butler \etal 2004) decays are
parallel.

\begin{figure}[h]
  \includegraphics[height=.36\textheight]{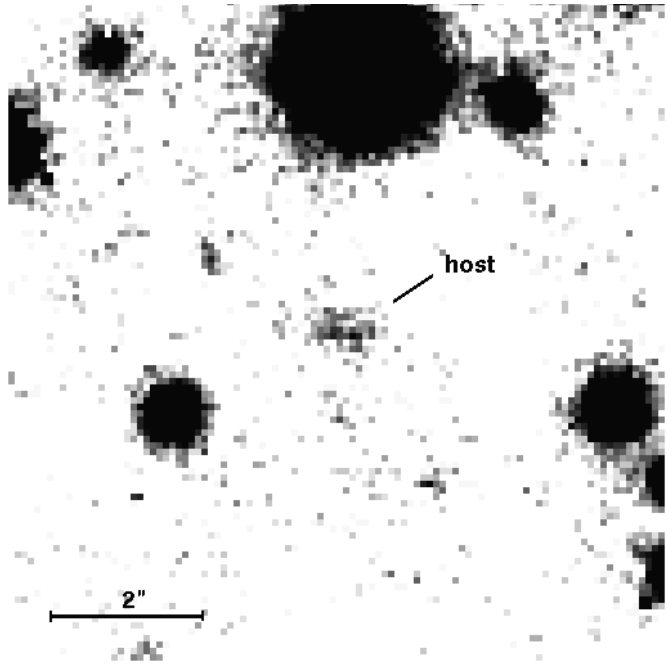}
  \caption{116\,min exposure $K$-band image taken with the 3.8\,m
  UKIRT equipped with the UFTI on June 12, 14.9\,days after the prompt
  emission. North is up and East to the left. The potential host
  galaxy (RA(2000)=17h04m00.3s, DEC(2000)=--22\degs37\amin10\asec of
  GRB 030528 shows an elongation in East-West direction.}
  \label{fig:Khost}
\end{figure}

The $K$-band image from an UKIRT/UFTI observation 14.9\,days after the
burst shows that the image of the source is extended relative to the
point spread function of the field in East-West direction
($\sim$1.5\asec$\times$0.8\asec) (Fig.~\ref{fig:Khost}). This
elongation is consistently seen in the late time ISAAC $J_s$-band
images. The center of the source is located at RA(2000)=17h04m00.3s,
DEC(2000)=--22\degs37\amin10s. The positional coincidence with the
afterglow suggests that this is the underlying host galaxy of GRB
030528. We cannot exclude a residual point-like contribution to the
total flux, including the afterglow and possibly an additional
supernova (Zeh \etal 2004).  A free fit to the $K$-band afterglow
light curve gives a host magnitude of $K$=19.9$\pm$0.7. Assuming that
after about 10 days all fluxes at shorter wavelengths ($V$--$J$) are
exclusively due to the host galaxy, we find a host magnitude of
$J$=20.8$\pm$0.1, an average $I$-band magnitude of $I$=21.3$\pm$0.3,
$R$=22.0$\pm$0.2 and $V$=21.9$\pm$0.2. In the $H$-band we use the
early SofI measurement of $H$=20.3$\pm$0.4 as an upper limit to the
host brightness.

These broadband photometric measurements can be used to constrain the
redshift range and galaxy classification of the host. Therefore, we
applied the photometric redshift technique described in Bender \etal
(2001). Thirty template spectra of a variety of galaxy types (from
ellipticals to late type spirals as well as irregulars) were convolved
with the filter curves and efficiencies used in our observations. The
templates consist of local galaxy spectra (Mannucci \etal 2001; Kinney
\etal 1996) and semi-empirical templates (Maraston 1998; Bruzual \&
Charlot 1993) and cover a wide range of ages and star formation
histories. Varying the redshift between $z$=0 and $z$=10 we determine
a probability density function via Bayesian statistic using
eigenspectra. This photometric redshift method was earlier applied and
calibrated e.g. on more than 500 spectroscopic redshifts of galaxies
from the Munich Near-Infrared Cluster Survey with an rms scatter of
0.055 and no mean bias (Drory \etal 2003).

%------------------------------------------------------------------
%------------------------------------------------------------------

\begin{figure}[h]
   \centering 
   \includegraphics[width=0.52\textwidth,angle=0]{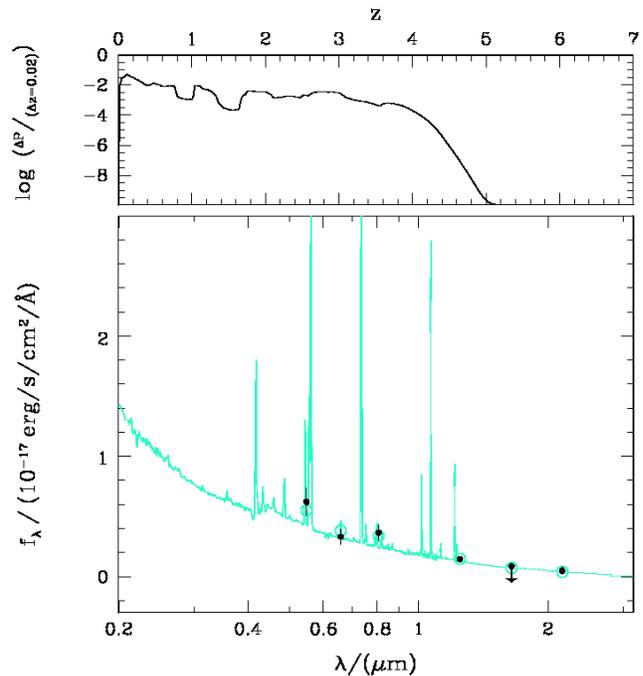}
   \caption{Results of the photometric redshift fit of C1 with
   template spectra following Bender \etal (2001). Upper panel: The
   probability density of the redshift of the object for late type
   galaxies. The probability for the redshift drops above $z\sim$4 and
   has several maxima below. Lower panel: The photometric measurements
   (points) together with the best fitting late type galaxy at a
   redshift of $z$=0.2 (solid line). Empty circles correspond to the
   expected magnitudes in the observed photometric bands.}
    \label{fig:hostProbLat} 
   \end{figure}

\begin{figure}[h]
   \centering 
   \includegraphics[width=0.52\textwidth,angle=0]{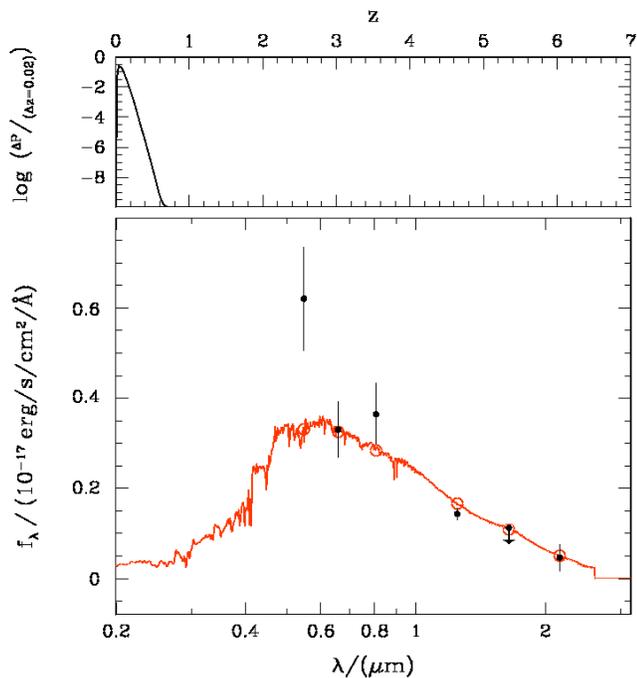}
   \caption{Same as Fig.~\ref{fig:hostProbLat} for early type galaxies.}
  \label{fig:hostProbEll} 
   \end{figure}

Fig.~\ref{fig:hostProbLat} and Fig.~\ref{fig:hostProbEll} show the
resulting redshift probability density function together with the
photometric measurements of the galaxy and two fitting galaxy template
spectra. Late type star forming galaxies lead to significantly better
matches to the observations (reduced $\chi$$^2$$<$1) in contrast to
ellipticals (reduced $\chi$$^2$$>$2). The lack of $U$-band
observations for GRB 030528 and the poor quality of the photometry
reduces the power of the method and results in a relatively
unconstrained redshift range. For late type galaxies redshifts beyond
of $z$=4 appeared to be ruled out, while for ellipticals redshifts
should not exceed $z=0.2$.  For completeness we also considered
stellar spectra, and find that all available templates produce fits
worse than those for elliptical galaxies (reduced $\chi^{2}$$>$2.2;
not plotted here).

%------------------------------------------------------------------
%------------------------------------------------------------------

%------------------------------------------------------------------
%------------------------------------------------------------------

\section{Discussion}

The discovery of the near-IR afterglow and optical/near-IR host galaxy
of GRB 030528 presented here, again demonstrates the importance of
rapid, deep multi-wavelength follow-up observations. In this
particular case rapid and deep near-IR observations were obtained, but
the afterglow would probably not have been identified without the {\it
Chandra} observations in the X-ray band. Guided by the X-ray
data, we discovered the afterglow in a crowded field at a position
significantly affected by image distortions. However, it took
additional X-ray observations to confirm the near-IR-candidate. The
chain of events for GRB 030528 emphasizes that observational programs
directed at the identification and study of GRB afterglows depend
critically on at least three ingredients; rapid response, deep
imaging, and multi wavelength coverage. Lacking any one of these, GRB
030528 would have most likely been labeled as a ``dark burst''. The
fact that it was caught in the near-IR still attaches the label
``optically dark'' GRB, and one wonders if this is an instrumental
effect or intrinsic.

The non-detection of the optical afterglow may have several
reasons. The lack of early, deep optical observations is perhaps the
most important. Using the decay observed in the $K_s$-band
(Fig.~\ref{fig:lightcurves}) we estimate the $R$-band magnitude at the
time of the early observations at the Bisei Observatory (Ayani \&
Yamaoka 2003) which provide a upper limit of $R$$>$18.7 at
t=0.097\,days. Ignoring the possibility of a break in the power law
decay one predicts $K_s\sim$16.5 at that time. With the typically
observed afterglow colors of $R$--$K$$\sim$2--5\,mag (Gorosabel \etal
2002) this corresponds to $R$=18.5--21.5. Applying the Galactic
foreground extinction correction the observable brightness should have
been $R$=20.1--23.1. An early break in the light curve would suggest
an even dimmer source. Clearly the depths of the Bisei observation did
not reach a level required to capture the afterglow of GRB
030528. Depending on the $R$--$K$ colour, the afterglow might
have been detected if it would not had suffered from the large
Galactic foreground extinction. Thus, the position in the Galactic
Plane additionally hampered the optical detection. It is obvious that
this GRB is yet another example of a burst which maybe ``falsely
accused of being dark''. {\it HETE-2} observations indeed suggest that
``optically dark'' burst may mostly be due to a shortage of deep
and rapid ground-based follow-up observations and adverse observing
conditions (Lamb \etal 2004).

A further challenge to the task of finding afterglows is the
competition in brightness between the afterglow and the host
galaxy. Except for extremely early observations the afterglow flux may
be comparable or significantly less than the integrated flux from a
normal galaxy. This obviously poses a challenge as the signal-to-noise
ratio is essential for any detection algorithm. 

\begin{figure}[h]
\centering \includegraphics[width=0.355\textwidth,angle=-90]{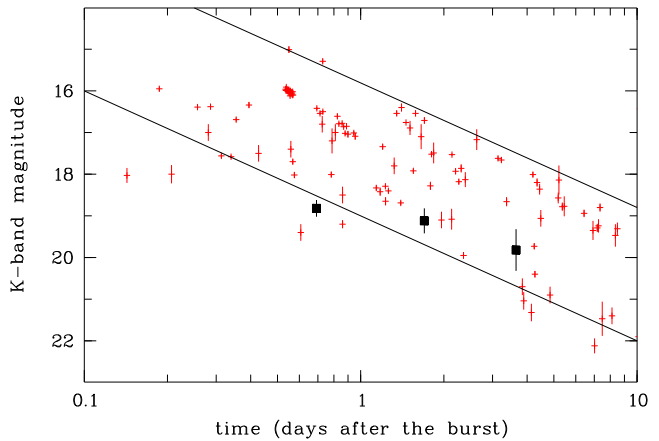}
\caption{Published $K$-band magnitudes of all GRB afterglows observed
  by May 2004 (dots; references listed in Appendix B) in
  comparison to the faint $K$-band afterglow of GRB 030528 (filled
  squares). Data are not corrected for Galactic extinction. The
  straight lines indicate a decay with a slope $\alpha=1.2$.  }
\label{fig:Kall}
\end{figure}

Considering the luminosity functions for the host galaxies and the
afterglows one potentially encounters a further challenge. A bright
afterglow against the backdrop of a faint host is probably easy to
identify, while a faint afterglow from a bright host could more easily
escape detection. Figure~\ref{fig:Kall} demonstrates that the $K$-band
afterglow of GRB 030528 was indeed very faint when compared to all
$K$-band detections of afterglows reported in the literature so
far. Only the $K$-band afterglow of \object{GRB 971214} was fainter at the time
of its discovery (Ramaprakash et al. 1998). On the other hand,
Fig.~\ref{fig:Kall} also shows that most $K$-band afterglows detected
by mid 2004 occupy a region, which spans over only 3
magnitudes. Assuming that $K$$\sim$19.6 measured at t=14.9\,days is in
fact the magnitude of the host (an underlying supernova contribution
might introduce a small upwards correction) this galaxy would be among
the brightest GRB hosts in the $K_s$-band to date.

% --------------------------------------------------------------------------
% --------------------------------------------------------------------------

Based on the few photometric data points for the afterglow, some of
which only provide upper limits due to the contribution of the host,
we derive a rough lower limit on the spectral slope, $\beta>0.4$. Such
a spectral slope would not be atypical. Also, the temporal decay is in
the range of the standard value of $\alpha=1.3$ (van Paradijs \etal
2000).

The near-IR afterglow was significantly detected only in the
$K_s$-band. Does the non-detection in the $J$ and $H$-bands require
large intrinsic extinction? Both, the $J$-band and the $H$-band
addresses the above question to some extent. Assuming a typical
afterglow law for the spectral energy distribution of
$F_\nu\propto\nu^{-\beta}$ with $\beta$=0.8 (0.4) we estimate
foreground extinction corrected $J$ and $H$-band magnitudes at
$t$=0.6\,days post burst. The expected $J$-band magnitude of
$J$$\sim$20.4 ($\sim$20.6) is consistent with the observed value of
$J$=20.3$\pm$0.4. As the host seems to dominate the $J$-band emission
at that time, no constraint on the intrinsic extinction of the
afterglow can be set. However, the expected $H$-band magnitude of
$H$$\sim$19.5 ($\sim$19.6) is significantly brighter than the observed
value of $H$=20.6$\pm$0.3 and thus indicative of some additional
intrinsic extinction. Nevertheless, the data only allow us to derive a
rough estimate for the observer frame extinction of $A_V$$>$2. From
estimates of the effective neutral hydrogen column density in the
X-ray band (Butler \etal 2004) it is clear that intrinsic extinction
in the host galaxy may account for at most a few magnitudes in the
$R$-band. These two approaches yield comparable values, but the
uncertainties are large in either case.

The $R$-$K$$\sim$2.4\,mag color of the host seems to be consistent
with the colors in the sample of GRB host galaxies detected in the
$K$-band (leFloc'h \etal 2003). Our spectral template fitting (see
section~5) is consistent with the idea that GRB host galaxies are
actively star forming blue galaxies as emphasized by leFloc'h \etal
(2003). It is perhaps fair to assume that the host of GRB 030528 is
similar to the host sample discussed by these authors. leFloc'h \etal
find that GRB hosts appear to be sub-luminous at approximately 8\% of
L$_*$ in a Schechter distribution function. If, for simplicity, we
assume that the host of GRB 030528 has an absolute brightness of
exactly this value (corresponding M$_K$=--22.25) the $K$-band
magnitude derived from the fit to the afterglow light curve of
$K$=19.9$\pm$0.7 implies a redshift of $z\sim$0.4--0.6 (for currently
accepted cosmological parameters). Applying the pseudo redshift
indicator of Atteia (2003b) to the {\it HETE-2} data gives $z$'=0.36,
which is close to the above value. The assumed absolute magnitude is
uncertain by about $\pm$2\,mag and pseudo redshifts are also uncertain
to within a factor two to three. It is thus clear that the redshift of
the host of GRB 030528 is by no means established. However, it seems
reasonable to interpret the observations to imply redshifts of the
order of unity or less. At $z$=0.4 the angular extent of 1\farcs5 of
the host (see Fig.~\ref{fig:Khost}) implies a linear dimension of
$\sim$8.7\,kpc (assuming standard cosmology). This would suggest the
host to be a blue compact star forming galaxy.
 
\bigskip
In summary, we have demonstrated that observations in the near-IR hold
the promise to detect afterglows that escape in the optical band
because of possible reddening. Despite the success of the discovery
of the afterglow of GRB 030528 coverage was insufficient to establish
a well sampled light curve and to derive an accurate
redshift. However, our detection of the host galaxy provides indirect
evidence for a low redshift. Despite a significant allocation of
observing time at large aperture telescopes, the sampling of this
afterglow fell short of optimal coverage and depth. The well
recognized shortage of global resources is likely to present a major
hurdle to afterglow programs in the {\it SWIFT}-era when the burst
detection rate is expected to increase dramatically.

% --------------------------------------------------------------------------
\section{Acknowledgements}

This work is primarily based on observations collected at the European
Southern Observatory, Chile, under the GRACE proposal 71.D-0355 (PI:
E.v.d.Heuvel) with additional data obtained at the Cerro Tololo
Inter-American Observatory and the United Kingdom Infra-Red
Telescope. We are highly indebted to the ESO staff, in particular
C. Cid, C. Foellmi, P. Gandhi, S. Hubrig, R. Johnson, R. Mendez,
J. Pritchard, L. Vanzi \& J. Willis for the prompt execution of the
observing requests and all additional effort related to that. We thank
the anonymous referee for insightful and helpful comments and P. Reig
for his attempt to observe the host with the 1.3\,m telescope at
Skinakas Observatory. This publication makes use of data products from
the Two Micron All Sky Survey, which is a joint project of the
University of Massachusetts and the Infrared Processing and Analysis
Center/California Institute of Technology, funded by the National
Aeronautics and Space Administration and the National Science
Foundation.  This research has made use of data obtained from the HETE
science team via the website
http://space.mit.edu/HETE/Bursts/Data. HETE is an international
mission of the NASA Explorer program, run by the Massachusetts
Institute of Technology.
\bigskip

\noindent{\bf Appendix A - CXOU J170354.0--223654}

\begin{figure}[h] \centering
   \includegraphics[width=0.5\textwidth,angle=0]{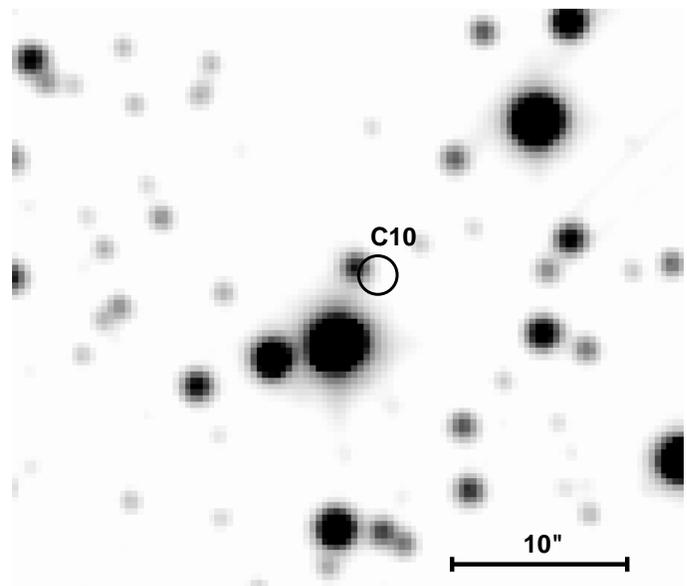}
   \caption{V-band image of the field around C10 taken with the
   D1m54 in June 2004. Coincident with the {\it Chandra} position
   (error circle as given in Butler \etal 2003a) we detect a point
   like source. North is up and East to the left.}
   \label{fig:chandra10Fc} 
   \end{figure}

We here summarize our results on the optical and near-IR observations
of C10=CXOU J170354.0--223654. At the X-ray position a source is
detected in the SofI $J$, $H$ and $K_s$ as well as in the Mosaic
$I$-band and DFOSC $V$ and $R$-band images
(Fig.~\ref{fig:chandra10Fc}, RA(2000)=17:03:54.0,
DEC(2000)=-22:36:53). C10 is outside the field of view of the ISAAC
and UFTI observations. Tab.~\ref{tab:CXOU10} lists the brightness and
flux density values for all observations with C10 in the field of
view. The source shows no significant variation in any band. Likewise,
the two {\it Chandra} observation also indicate that C10 is constant
in the X-ray band. This makes an association with GRB 030528 very
unlikely.

%------------------------------------------------------------------
%------------------------------------------------------------------

%------------------------------------------------------------------
\begin{table}[h]
\begin{center}
\caption{Brightness and flux densities of C10 corrected for foreground extinction in the Galaxy (assuming an extragalactic origin).}
\begin{tabular}{cccc}
\hline\hline
$<dt>$ & Filter & Brightness & log Flux\\
(days) & & (mag) & (erg/cm$^2$/s/\AA) \\
\hline
0.6684 & J & 16.04$\pm$0.05 & --15.91$\pm$0.02\\
1.6673 & J & 16.20$\pm$0.24 & --15.97$\pm$0.09\\
0.6809 & H & 15.78$\pm$0.05 & --16.24$\pm$0.02 \\
1.6826 & H & 15.86$\pm$0.08 & --16.27$\pm$0.03\\
0.6920 & K$_s$ & 15.69$\pm$0.06 & --16.67$\pm$0.02\\
1.6993 & K$_s$ & 15.76$\pm$0.10 & --16.69$\pm$0.04\\
3.6486 & K$_s$ & 15.62$\pm$0.06 & --16.64$\pm$0.02\\
6.5694 & I  & 16.42$\pm$0.05 & --15.80$\pm$0.02\\
32.5951 & I  & 16.36$\pm$0.05 & --15.77$\pm$0.02\\
382.5 & R & 16.81$\pm$0.05 & --15.47$\pm$0.02\\
384.5 & V & 17.07$\pm$0.05 & --15.26$\pm$0.02\\
\hline\hline
\end{tabular}
\label{tab:CXOU10}
\end{center}
\end{table}
%------------------------------------------------------------------

\begin{figure}[h]
   \centering
   \includegraphics[width=0.52\textwidth,angle=0]{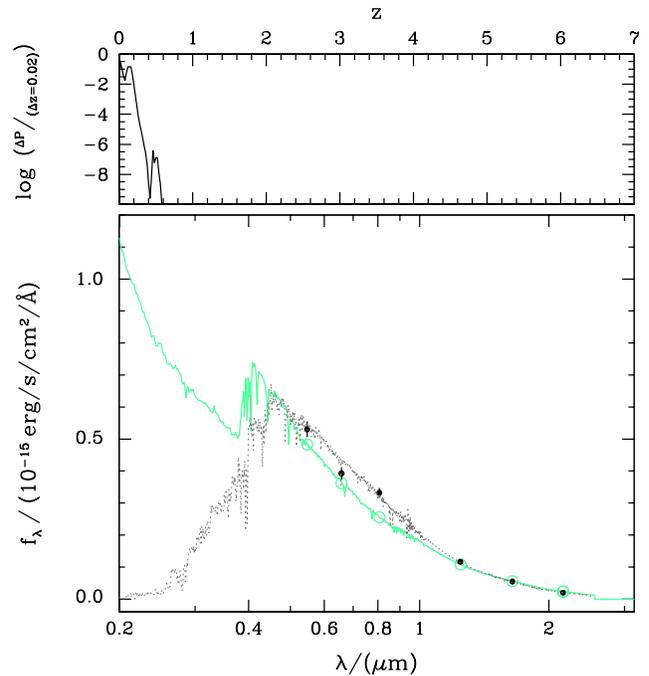} 
   \caption{Same as Fig.~\ref{fig:hostProbLat} for C10. Upper panel:
   A global maximum of the probability density is evident at
   $z<$0.2. Lower panel: Best fitting (late type) galaxy model
   (solid line) and best fitting (G0) stellar spectrum (dotted line).}
   \label{fig:chandra10Prob} 
   \end{figure}

The counterpart of C10 appears point like in the optical/near-IR
images (Fig.~\ref{fig:chandra10Fc}). This allows both, a stellar and a
Galactic nature of the object. In case of a star, the distance of the
source in the Galaxy and thus the degree of extinction is
unknown. Exploring the range from a nearby location (quasi
unextinguished) to the opposite side of the Milky Way (foreground
extinction assumed as given above) we derive ranges for the near-IR
colors of $J$-$K$=0.2--0.6\,mag and $I$-$K$=0.6-1.6\,mag. Due to the
unknown extinction, the optical colours span an even wider range. The colors
constrain the spectral type to G--K for a main sequence star or
supergiant (Johnson 1966). Fig.~\ref{fig:chandra10Prob} shows an
example fit of the spectrum of a G0 star to our photometric data
(lower panel; dotted line). However, the ratio of X-ray to bolometric
flux can be used to test the stellar origin of C10. From the observed
{\it Chandra} counts in the energy range of 0.5--8\,keV (a total of
17.5 counts in 45\,ksec) we derive the X-ray flux as
F$_x$$\sim$(3.3--4.2)$\times$10$^{-15}$ erg cm$^{-2}$ s$^{-1}$
(assuming a power law shape with $\Gamma$=1.9). The lower value gives
the flux corrected for the hydrogen column in the Galaxy
(N$_H$=1.6$\times$10$^{21}$\,cm$^{-2}$) while the upper limit results
for an object with N$_H$=0. Using the colors and bolometric
corrections for G--K main sequence stars given by Johnson (1966)
together with the observed $K$-band magnitudes and filter width, we
estimate the bolometric flux to be F$_{bol}$=(0.8--4)$\times$10$^{-13}$
erg cm$^{-2}$ s$^{-1}$. The resulting ratio of
log(F$_x$/F$_{bol}$)=-1.8$\pm$0.5 exceeds significantly the value
typically observed for G--K stars (-4.5 to -6.5; Pallavicini \etal
1981). Therefore, although the photometric data allow an association
of C10 with a single star, the X-ray properties make this very
unlikely. Nevertheless, we can not exclude an X-ray binary system in
the Galactic Plane with the X-ray radiation coming from an accretion
disk around a black hole or from a neutron star surface and the
optical/near-IR emission being produced by the accretion disk, outflow
and secondary star.

%------------------------------------------------------------------
%------------------------------------------------------------------

C10 can also be associated with a galaxy behind the Milky
Way. Similarly to the analysis of C1, we applied the photometric
redshift technique of Bender \etal
(2001). Fig.~\ref{fig:chandra10Prob} shows the redshift probability
density function and the best fitting galaxy spectrum. A fit of a late
type galaxy template to the data shows no sufficient match (reduced
$\chi$$^{2}\sim$2.3). Early type galaxies are excluded by the fit with
a reduced $\chi$$^{2}$$>$6. The redshift for the best fitting galaxy
template is $z<$0.2. Given the point-like appearance of the
source, $z$=0.1 would require a compact small galaxy. As for a star,
we use the X-ray properties to test for a possible galaxy nature of
C10. From the best fitting galaxy template spectrum the expected
$B$-band flux can be estimated. A late type galaxy with the observed
multicolor magnitudes as shown in Fig.~\ref{fig:chandra10Prob}, has
F$_B$$\sim$3.6$\times$10$^{-13}$\,erg cm$^{-2}$ s$^{-1}$. Using the
extinction corrected X-ray flux given above, we derive
log(F$_x$/F$_{B}$)=--2.6 consistent with the observations of normal
galaxies (Fabbiano 1989).

The applied photometric redshift method does not allow us to estimate
the redshift probability for AGNs. As shown above, templates of stars
and normal galaxies have problems to fit the data. The appearance as a
point source together with the X-ray properties instead make an
association of C10 with an AGN the most likely solution. Spectroscopic
observations are necessary to confirm this result.
\bigskip

\noindent{\bf Appendix B - $K$-band afterglow light curve data}

We used the following bursts for the compilation
of $K$-band afterglow light curves in Fig.\ref{fig:Kall}:
 GRB 970508 (Chary \etal 1998),
 GRB 971214 (Gorosabel \etal 1998; Ramaprakash \etal 1998),
 GRB 980329 (Larkin \etal 1998; Reichart \etal 1999),
 GRB 980613 (Hjorth \etal 2002),
 GRB 990123 (Kulkarni \etal 1999; Holland \etal 2004),
 GRB 991208 (Bloom \etal 1999), 
 GRB 991216 (Vreeswijk \etal 1999; Garnavich \etal 2000; Halpern \etal 2000),
 GRB 000131 (Andersen \etal 2000),
 GRB 000301C (Kobayashi \etal 2000; Jensen \etal 2001; Rhoads \& Fruchter 2001),
 GRB 010222 (Masetti \etal 2001),
 GRB 001011 (Gorosabel \etal 2002),
 GRB 000926 (Di Paola \etal 2000; Fynbo \etal 2001b),
 GRB 011121 (Price \etal 2002; Greiner \etal 2003b),
 GRB 011211 (Jakobsson \etal 2003),
 GRB 020305 (Burud \etal 2002),
 GRB 020322 (Mannucci \etal 2002),
 GRB 020405 (Masetti \etal 2003),
 GRB 020813 (Covino \etal 2003),
 GRB 021004 (Di Paola \etal 2002),
 GRB 021211 (Fox \etal 2003b),
 GRB 030115 (Kato \& Nagata 2003; Dullighan \etal 2004),
 GRB 030227 (Castro-Tirado \etal 2003),
 GRB 030323 (Vreeswijk \etal 2004),
 GRB 030429 (Nishiyama \etal 2003; Jakobsson \etal 2004),
 XRF 030723 (Fox \etal 2003a; Fynbo \etal 2004),
 GRB 031203 (Malesani \etal 2004; Prochaska \etal 2004).

% --------------------------------------------------------------------------

\end{document}